\documentclass{bioinfo}
\copyrightyear{2005}
\pubyear{2008}

\newcommand{\go}{G_{0}}
\newcommand{\gl}{G_{1}}
\newcommand{\gtn}{G_{29}}

\begin{document}
\firstpage{1}

\title[Graphlet Degree Signatures and Biological Function]{Uncovering Biological Network Function via Graphlet Degree Signatures}
\author[{Milenkovi\'{c} and Pr\v{z}ulj}]{Tijana Milenkovi\'{c}\, and Nata\v{s}a Pr\v{z}ulj\,\footnote{to whom correspondence should be addressed}}
\address{Department of Computer Science, University of California, Irvine, CA 92697-3435, USA\\
\bigskip
{\bf{Technical Report No. 08-01, Donald Bren School of Information and Computer Sciences, University of California, Irvine, USA, February 2008.}}\\
\medskip
\small{First submitted to Nature Biotechnology on July 16, 2007.\\
\medskip
Presented at BioPathways'07 pre-conference of ISMB/ECCB'07, July 19-20, 2007, Vienna, Austria. \\
\medskip
Published in full in the Posters section of the Schedule of the RECOMB Satellite Conference on Systems Biology, November 30 - December 1, 2007, University of California, San Diego, USA.\\
}
}

\history{}

\editor{}

\maketitle

\begin{abstract}

\section{Motivation:}
Proteins are essential macromolecules of life and thus understanding
their function is of great importance.  The number of functionally
unclassified proteins is large even for simple and well studied
organisms such as baker's yeast.  Methods for determining protein
function have shifted their focus from targeting specific proteins
based solely on sequence homology to analyses of the entire proteome
based on protein-protein interaction (PPI) networks.  Since proteins
aggregate to perform a certain function, analyzing structural
properties of PPI networks may provide useful clues about the
biological function of individual proteins, protein complexes they
participate in, and even larger subcellular machines.

\section{Results:}
We design a sensitive graph theoretic method for comparing local
structures of node neighborhoods that demonstrates that in PPI
networks, biological function of a node and its local network
structure are closely related.  The method groups topologically
similar proteins under this measure in a PPI network and shows that
these protein groups belong to the same protein complexes, perform the
same biological functions, are localized in the same subcellular
compartments, and have the same tissue expressions.  Moreover, we
apply our technique on a proteome-scale network data and infer
biological function of yet unclassified proteins demonstrating that
our method can provide valuable guidelines for future experimental
research.

\section{Availability:}
Data is available upon request.

\section{Contact:} \href{tmilenko@ics.uci.edu}{natasha@ics.uci.edu}
\end{abstract}

\vspace{-0.1cm}

\section{Introduction}\label{introduction}

Large amounts of biological network data are becoming available.  
We study \emph{protein-protein interaction (PPI) networks} 
(or \emph{graphs}), in which
nodes correspond to proteins and undirected edges represent
physical interactions between them.
Since a protein almost never 
acts in isolation, but
rather interacts with other proteins in order to perform a certain
function, PPI networks by
definition reflect the interconnected nature of biological processes.
Analyses of PPI networks may give
valuable insight into biological mechanisms and provide deeper
understanding of complex diseases. 
Defining the relationship between the PPI network topology and biological function and
inferring protein function from it is one of the major challenges in the post-genomic era \citep{Nabieva2005,Vazquez2003,Schwikowski2000,Hishigaki2001,Letovsky2003,Deng2003,Deng2004,Brun2004}.

\vspace{-0.2cm}
\subsection{Background}\label{background}

Various approaches for determining protein function from PPI networks
have been proposed.  ``Neighborhood-oriented'' approaches observe the
neighborhood of a protein to predict its function by finding the most
common function(s) among its neighbors. The ``majority rule'' approach
considers only nodes directly connected to the protein of interest
\citep{Schwikowski2000}.  An improvement is made by also observing
indirectly connected level-2 neighbors of a node
\citep{Chua2006}.  Furthermore, the 
function with the highest $\chi$$^2$ value amongst
the functions of all ``$n$-neighboring proteins'' is assigned to the
protein of interest \citep{Hishigaki2001}.  
Other approaches use the idea of shared neighbors \citep{Samanta2003} or 
the network flow-based idea \citep{Nabieva2005} to determine protein function. 

Several global optimization-based function prediction strategies have
also been proposed.  Any given assignment of functions to the whole
set of unclassified proteins in a network is given a score, counting
the number of interacting pairs of nodes with no common function; the
functional assignment with the lowest score maximizes the presence of
the same function among interacting proteins \citep{Vazquez2003}.  An
approach that reduces the computation requirements of this method has
been proposed \citep{Sun2006}.

Cluster-based approaches are exploiting the existence of regions in
PPI networks that contain a large number of connections between the
constituent proteins.  These dense regions are a sign of the common
involvement of those proteins in certain biological processes and
therefore are feasible candidates for biological complexes.  The
restricted-neighborhood-search clustering algorithm efficiently
partitions a PPI network into clusters identifying known and
predicting unknown protein complexes \citep{King04}.  Similarly, highly
connected subgraphs are used to identify clusters in networks
\citep{Sham1}, defining the relationship between the PPI network size and
the number and complexity of the identified clusters, and identifying
known protein complexes from these clusters \citep{Przulj03}.  
Moreover, Czekanowski-Dice distance is used for protein function 
prediction by forming clusters of proteins sharing a high
percentage of interactions \citep{Brun2004}.

\subsection{Approach}\label{our_approach}

We address the above mentioned challenge. First, we verify that in PPI
networks of yeast and human, local network structure and biological
function are closely related.  We do this by designing a method that
clusters together nodes of a PPI network with similar topological
surroundings and by demonstrating that it successfully uncovers groups
of proteins belonging to the same protein complexes, performing the
same biological functions,
being localized in the same subcellular compartments, and having the
same tissue expressions.  Since we verify this for PPI networks of a
unicellular and a multicellular eukaryotic organism (yeast and human,
respectively), we hypothesize that PPI network structure and biological
function are related in other eukaryotic organisms as well.
Next, since the number of functionally unclassified proteins is large
even for simple and well studied organisms such as baker's yeast
\emph{Saccharomyces cerevisiae}
\citep{PenaCastillo2007}, we describe how to apply our technique to
predict membership in protein complexes, functional groups, and
subcellular compartments of yet unclassified yeast proteins.

Our method belongs to the group of clustering-based approaches. 
However, compared to other 
methods that define a cluster as a dense interconnected region of a
network, our method defines it as a set of nodes with similar
topological \emph{signatures} (defined below).  Thus, nodes belonging to
the same cluster do not need to be connected or belong to the same
part of the network.

\begin{methods}
\section{Methods}\label{methods}

Our new measure of node similarity 
generalizes the degree of a node, which counts the
number of edges that the node touches, into the vector of
\emph{graphlet degrees}, counting the number of graphlets that the node
touches; \emph{graphlets} are small connected non-isomorphic induced
subgraphs of a large network \citep{Przulj04} (see
Figure \ref{fig:graphlet_orbits}).  As opposed to \emph{partial}
subgraphs (e.g., network \emph{motifs} \citep{Milo02}), graphlets must
be \emph{induced}, i.e., they must contain all edges between the nodes
of the subgraph that are present in the large network.  We count the
number of graphlets touching a node for all 2-5-node graphlets,
denoted by $\go$, $\gl$, $\ldots$, $\gtn$ in
Figure \ref{fig:graphlet_orbits}; counts involving larger graphlets
become computationally infeasible for large networks. Clearly, the
degree of a node is the first one in this vector, since an edge
(graphlet $G_0$) is the only 2-node graphlet.  
We call this vector the
\emph{signature} of a node.  
It is topologically relevant to distinguish between nodes touching a 3-node linear path
(graphlet G$_1$) at an end, or at the middle node; we provide a
mathematical formulation of this phenomenon for all graphlets with 2-5
nodes.  This is summarized by
\emph{automorphism orbits} (or just
\emph{orbits}, for brevity): by taking into account the ``symmetries''
between nodes of a graphlet, there are 73 different orbits for
2-5-node graphlets, numerated from 0 to 72 in Figure \ref{fig:graphlet_orbits}
(see \citep{Przulj2006} for details).  Thus, the signature vector
of a node has 73 coordinates.

\begin{figure}[ht]
        \centering
        \scalebox{.35}{\includegraphics{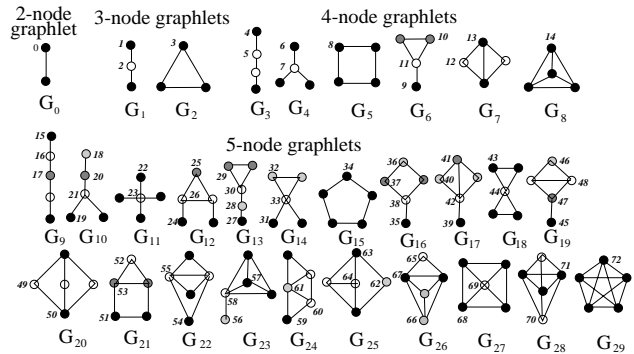}}
        \caption[The Graphlet automorphism orbits]{{The thirty 2-, 3-, 4-, and 5-node graphlets $G_0, G_1, \ldots, G_{29}$ and their automorphism orbits $0,1,2, \ldots, 72$.  In a graphlet $G_i$, $i \in \{0,1,\ldots29\}$, nodes belonging to the same orbit are of the same shade \citep{Przulj2006}.}}
        \label{fig:graphlet_orbits}
\end{figure}

We compute node signature similarities as follows.
We define a 73-dimensional vector $W$ containing the weights
\emph{w$_{i}$} corresponding to orbits $i \in \{0, \ldots, 72\}$, 
where weights are determined as follows. For each orbit, we consider
the number of orbits affecting it.  For example, the differences
in orbit $0$ (i.e., in the degree) of two nodes will automatically imply
the differences in all other orbits, since all orbits depend on
it.  Each orbit $i$ is assigned an integer \emph{o$_{i}$} that
represents the number of orbits that affect it (available upon request). We consider that each
orbit affects itself. 
We compute $w_i$ as a function of $o_i$.
We need to assign a higher weight $w_i$ to
the orbits that are not affected by many other orbits.  Thus, we apply
a slow-increasing logarithm function to $o_i$s.  Also, since the
maximum value that an $o_i$ can take is $73$ (for 2-5-node graphlets),
we divide $log_{10}(o_i)$ by $log_{10}(73)$ to scale it to [0, 1].
Since an orbit dependency count $o_i$ of 1 indicates that no other
orbits affect orbit $i$ (i.e., this orbit is of the highest importance),
we invert this scaled value of orbit dependencies as 
\begin{center}
\begin{equation*} 
w_i = 1 - \frac{log_{10}(o_i)}{log_{10}(73)}
\label{eq:01}	
\end{equation*}
\end{center}
to assign the highest weight of $1$
to orbit $i$ with $o_i=1$.  Clearly, $w_i \in [0,1]$ for all $i \in
\{0,\ldots,72\}$ and orbits become less important as their weights
$w_i$ decrease.  

For a node $u$, we denote by $u_i$ the $i^{th}$ coordinate
of its signature vector, i.e., $u_{i}$ is the number of times node $u$
touches orbit $i$. We define the distance $D_{i}(u,v)$ between
the $i^{th}$ orbits of nodes $u$ and $v$ as:
\begin{center}
\begin{equation*} 
D_i(u,v) = w_i \times \frac{|log_2(u_i + 1) - log_2(v_i + 1)|}{log_2(max\{u_i, v_i\} + 2)}.
\label{eq:02}	
\end{equation*}
\end{center}
We use $log_2$ in the numerator because the $i^{th}$ coordinates of
signature vectors of two nodes can differ by several orders of
magnitude and we do not want the distance measure to be entirely
dominated by these large values. Also, by using these logarithms, 
we take into account the relative difference between $u_i$ and $v_i$ instead of the absolute difference.  
We add 1 to $u_{i}$ and $v_{i}$ in the numerator of the formula for
$D_i(u,v)$ to prevent the logarithm function to go to infinity.  We
scale $D_{i}$ to be in [0, 1] by dividing with the value of the
denominator in the formula for $D_i(u,v)$.  We add 2 in the denominator of
the formula for $D_i(u,v)$ to prevent it from being infinite or 0.  We
find the total distance $D(u,v)$ between nodes $u$ and $v$ as:
\begin{center}
\begin{equation*} 
D(u,v) = \frac{\sum_{i=0}^{72}D_i}{\sum_{i=0}^{72}w_i}. 
\label{eq:03}	
\end{equation*}
\end{center}
Clearly, the distance $D(u,v)$ is in [0, 1], where distance 0 means
the identity of signatures of nodes $u$ and $v$.
Finally, the \emph{signature similarity}, $S(u,v)$, between nodes $u$ and $v$
is: 

\begin{center}
\begin{equation*} 
S(u,v) = 1 - D(u,v).
\label{eq:04}	
\end{equation*}
\end{center}

\vspace{0.2cm}

For a node of interest, we form a cluster containing that node and
all nodes in a network that are similar to it.  
According to the signature similarity metric, nodes $u$ and $v$ will be in the same
cluster if their signature similarity $S(u,v)$ is above a chosen
threshold.  
%
%
We choose an experimentally determined thresholds of 0.9-0.95. 
For thresholds above these values, only a few small clusters are
obtained, especially for smaller PPI networks, 
indicating too high stringency in signature similarities.
For thresholds bellow 0.9, the clusters are very large,
especially for larger PPI networks, 
indicating a loss of signature similarity.  
%
%
To illustrate signature similarities and our choices of signature
similarity thresholds, in Figure \ref{fig:signatures} we present the
signature vectors of yeast proteins in the PPI network of \citep{Krogan2006}
with signature similarities above 0.90
(Figure \ref{fig:signatures} A) and below 0.40
(Figure \ref{fig:signatures} B).  Signature vectors of proteins with
high signature similarities follow the same pattern, while those of
proteins with low signature similarities have very different patterns.

\begin{figure*}[hbtp]
\begin{center}
\begin{tabular}{cc}
\textsf{(A)} \resizebox{0.455\textwidth}{!}{\includegraphics{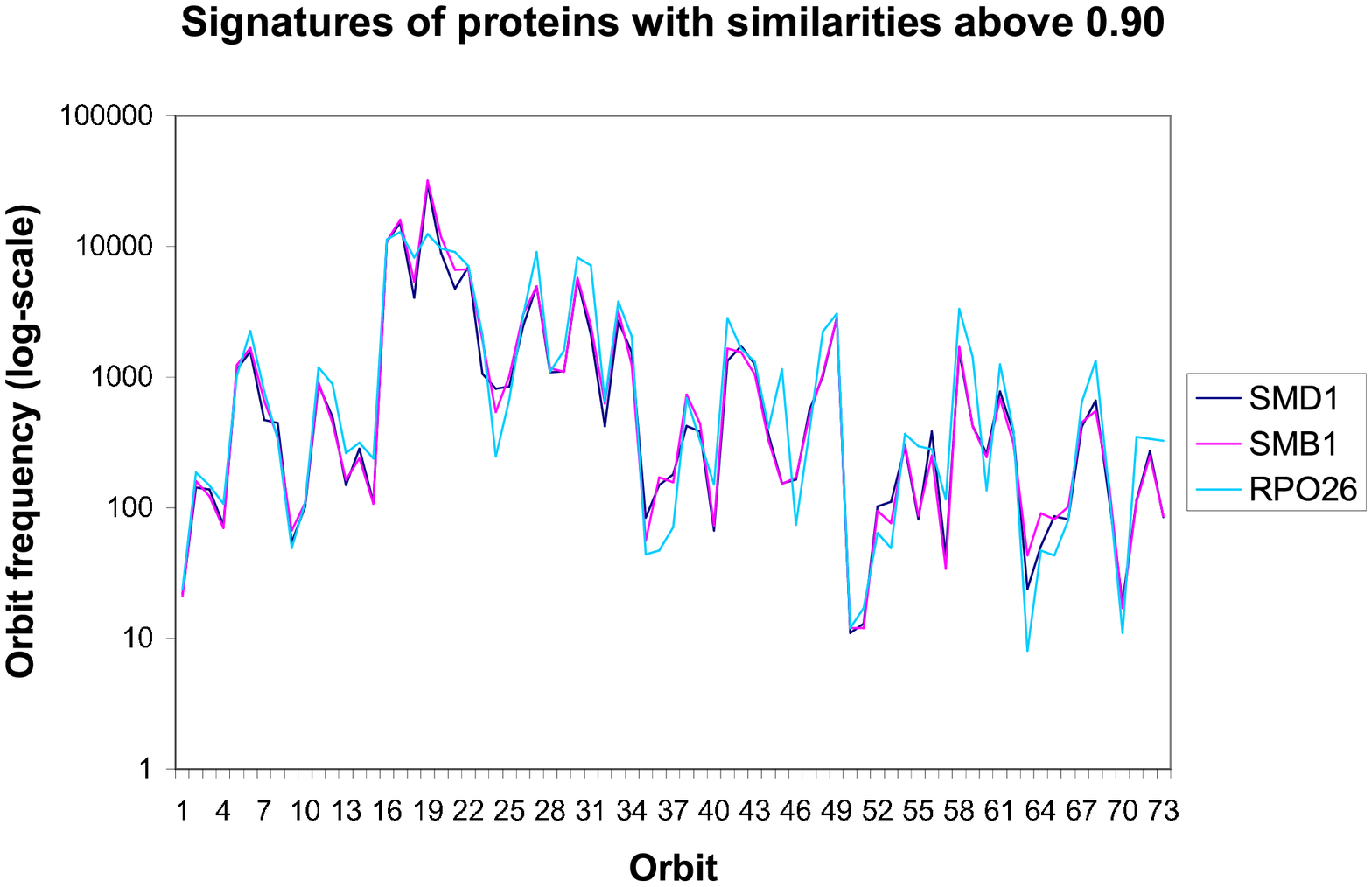}} &
\textsf{(B)} \resizebox{0.455\textwidth}{!}{\includegraphics{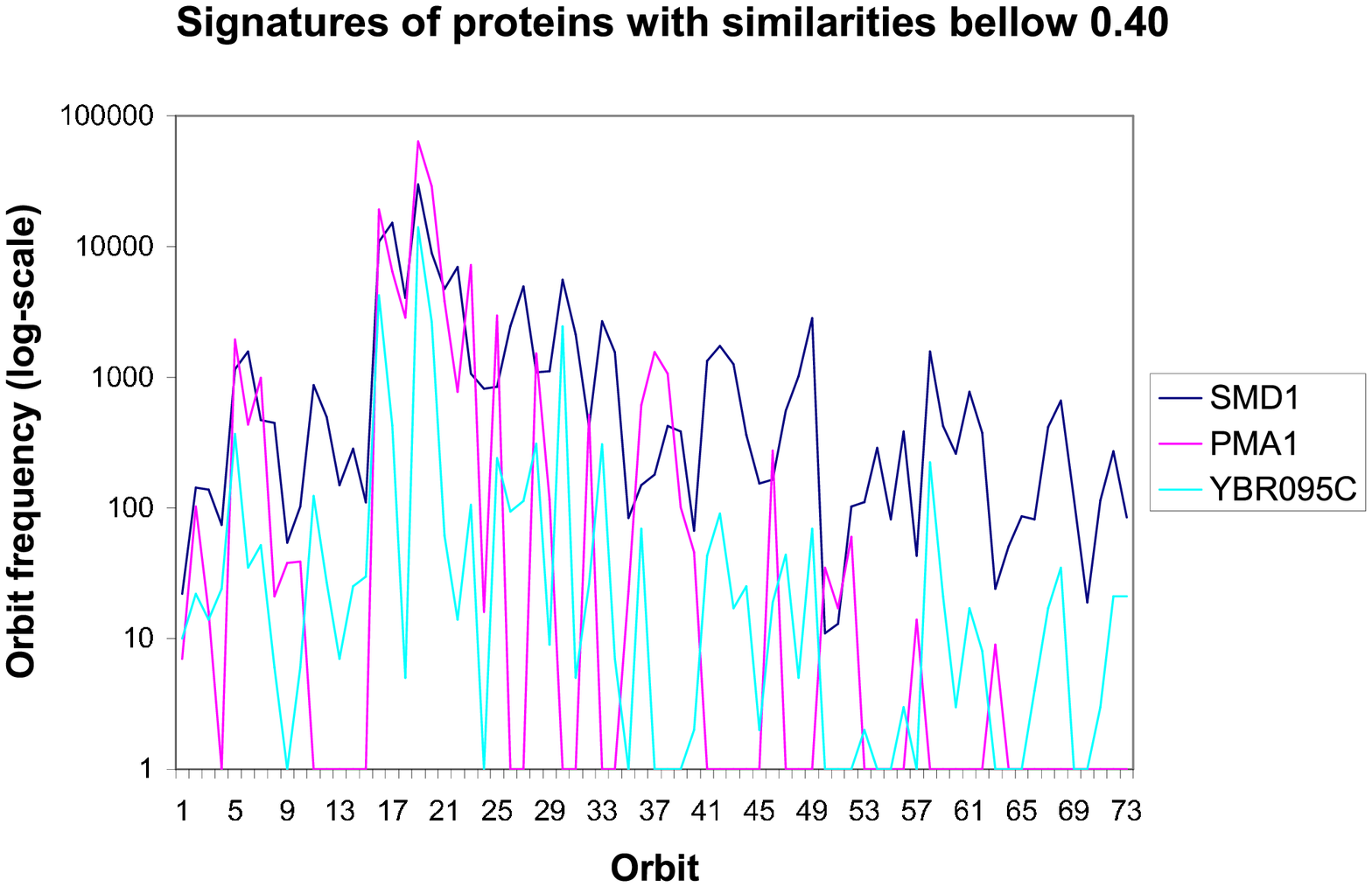}} 
\end{tabular}
\caption{
Signature vectors of proteins with signature similarities: (A) above
0.90; and (B) below 0.40.  The 73 orbits are presented on the abscissa
and the numbers of times that nodes touch a particular orbit are
presented on the ordinate in log scale.  In the interest of the
aesthetics of the plot, we added 1 to all orbit frequencies to avoid
the log-function to go to infinity in the case of orbit frequencies of
0.}
\label{fig:signatures}
\end{center}
\end{figure*}

\end{methods}

\section{Results and Discussion}\label{results}

We apply our method to six \emph{S. cerevisiae} PPI
networks and three \emph{human} PPI networks.  The \emph{S. cerevisiae}
PPI networks are henceforth denoted by ``vonMering-core'' \citep{Mering02},
``vonMering'' \citep{Mering02}, ``Krogan'' \citep{Krogan2006},
``DIP-core'' \citep{Deane2002}, ``DIP'' \citep{DIP}, and ``MIPS''
\citep{MIPS}.  
``vonMering-core'' contains only high-confidence
interactions described by von Mering et al. \citep{Mering02}; it
contains 2,455 interactions amongst 988 proteins obtained mainly by
tandem affinity purification (TAP) \citep{Rigaut99,Gavin02} and High-Throughput
Mass Spectromic Protein Complex Identification (HMS-PCI)
\citep{Ho2002}.  ``vonMering'' is the PPI network containing the top
11,000 high-, medium-, and low-confidence interactions amongst 2,401
proteins described by von Mering et al. \citep{Mering02}; the dominant
techniques used to identify PPIs in this network are TAP, HMS-PCI,
gene neighborhood, and yeast-two-hybrid (Y2H). ``Krogan'' is the
``core'' PPI data set containing 7,123 interactions amongst 2,708
proteins obtained by TAP experiments as described by Krogan et
al. \citep{Krogan2006}.  ``DIP-core'' is the more reliable subset of
the yeast PPI network from DIP
\citep{DIP} as described by Deane et al. \citep{Deane2002}; 
it contains 5,174 interactions amongst 2,210 proteins. ``DIP'' and
``MIPS'' are the yeast PPI networks downloaded in November 2007 from DIP
\citep{DIP} and MIPS \citep{MIPS} databases, respectively; they contain 
17,201 and 12,525 interactions amongst 4,932 and 4,786 proteins,
respectively.  
The three human PPI networks that we analyze are henceforth
denoted by ``BIOGRID'' \citep{BIOGRID}, ``HPRD'' \citep{HPRD}, and
``Rual'' \citep{Rual05}.  ``BIOGRID'' and ``HPRD'' are the human PPI
networks downloaded in November 2007 from ``BIOGRID'' \citep{BIOGRID}
and ``HPRD'' \citep{HPRD} databases, respectively; they contain 23,555
and 34,119 interactions amongst 7,941 and 9,182 proteins,
respectively.  ``Rual'' is the human PPI network containing 3,463
interactions amongst 1,873 proteins, as described by Rual et
al. \citep{Rual05}.
%
%
We removed all self-loops and multiple edges from each of the PPI networks 
that we analyzed.

The entire PPI network is taken into account when computing signature
similarities between pairs of nodes (i.e., proteins) and forming
clusters (see section \ref{methods}).  However, here we only report
the results of analyzing proteins involved in more than four
interactions.  We discard poorly connected proteins from our clusters
because they are more likely to be involved in noisy interactions.
Similar was done by Brun et al. \citep{Brun2004}.  Note that the
highest node degree in the analyzed PPI networks is 286.
Also, we discard very small clusters containing less than three proteins.  
%
%
For the remaining clusters, we search for common \emph{protein
properties}: in yeast PPI networks, we look for the common protein
complexes, functional groups, and subcellular localizations (described
in MIPS \citep{MIPS}) of proteins belonging to the same cluster; in
human PPI networks, we look for the common biological processes,
cellular components, and tissue expressions (described in HPRD
\citep{HPRD}) of proteins in the same cluster.

Classification schemes and the data for the three protein properties
that we analyzed in yeast PPI networks were downloaded from MIPS database
\citep{MIPS} in November 2007.  For each of these three classification
schemes (corresponding to protein complexes, biological functions, and
subcellular localizations), we define two levels of strictness: the
\emph{strict} scheme uses the most specific MIPS annotations, and the
\emph{flexible} one uses the least specific ones. For example, for a
protein complex ``category'' annotated by \emph{510.190.900} in MIPS, the
strict scheme returns
\emph{510.190.900}, and the flexible one returns \emph{510}.  
Classification schemes and the data for the three protein properties
that we analyzed in human PPI networks (corresponding to biological processes,
cellular components, and tissue expressions) were downloaded from HPRD database
\citep{HPRD} in November 2007.
In order to test if our method clusters together proteins having the
same protein properties, we refine our clusters by removing the nodes
that are not contained in any of the yeast MIPS protein complex,
biological function, or subcellular localization categories, or in any
of the human HPRD biological process, cellular component, or tissue expression
categories, respectively.
%

\begin{figure}
\centering
\resizebox{0.48\textwidth}{!}{\includegraphics{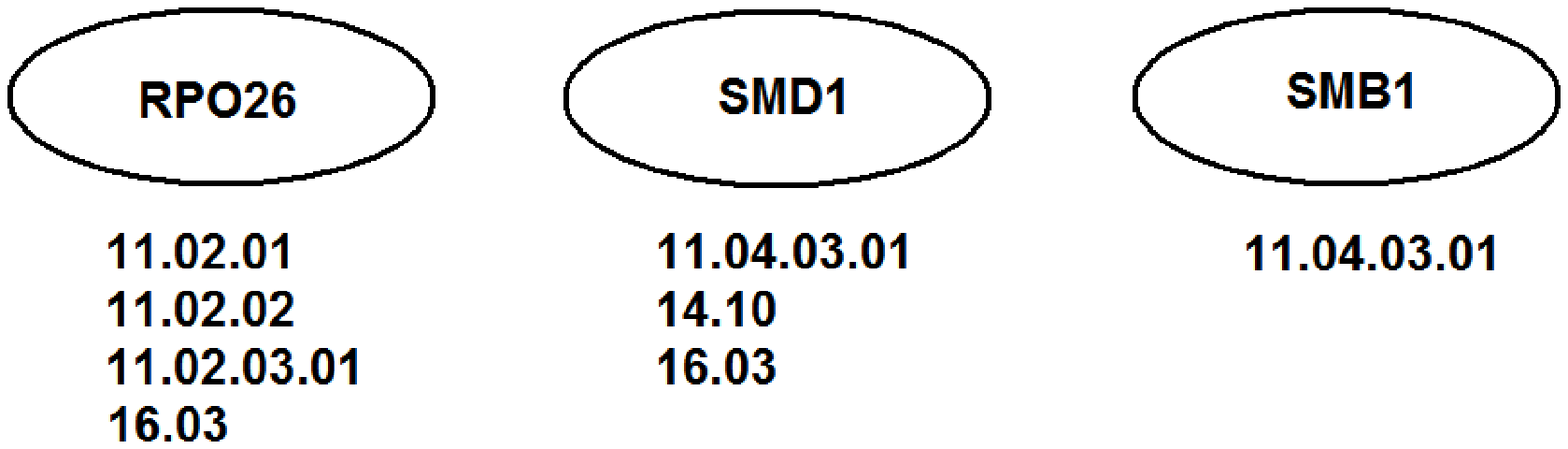}}
\caption{
An example of a three-node cluster, consisting of proteins RPO26,
SMD1, and SMB1.  The categories of biological functions that the
proteins belong to are presented bellow the protein names.}
\label{fig:example}
\end{figure}

In our clusters, we measure the size of the largest common category
for a given protein property as the percentage of the cluster size; we
refer to it as the \emph{hit-rate}.  Clearly, a yeast protein can
belong to more than one protein complex, be involved in more than one
biological function, or belong to more than one subcellular
compartment (and similar holds for human proteins).  Thus, it is possible to
have an overlap between categories, as well as more than one largest
category in a cluster for a given protein property.
%
%
We illustrate this for biological functions in the cluster presented
in Figure \ref{fig:example}, consisting of yeast proteins RPO26, SMD1,
and SMB1.  According to the strict scheme, protein SMD1 is in the
common biological function category with protein RPO26 (16.03), as
well as with protein SMB1 (11.04.03.01).  Thus, there are two largest
common biological function categories.  The size of the largest common
biological function category in the cluster is two and the hit-rate is
2/3=67\%.  For the flexible scheme, all three proteins are in one
common biological function category (11) and thus, the size of the largest
common biological function category is three and the hit-rate is 3/3=100\%.

We also define the \emph{miss-rate} as the percentage of the nodes in
a cluster that are not in any common category with other nodes in the
cluster, for a given protein property.  For example, in
Figure \ref{fig:example}, according to the strict scheme, proteins RPO26
and SMB1 are in a common biological function category with SMD1, but
they themselves are not in any common biological function category.  Although
not all three proteins are in the same biological function category and the
hit-rate is only 67\%, the miss-rate is 0/3=0\%, since every node is
in at least one common biological function category with another node in the
cluster.  Clearly, the miss-rate for the flexible scheme is also
0/3=0\%, since the three proteins are in the same biological function category
(11) with respect to this scheme.  Thus, if a protein belongs to
several different categories for a given protein property (which is
expected), the hit-rate in the cluster might be lower than 100\% (as
illustrated in Figure
\ref{fig:example}).  
Therefore, miss-rates are additional indicators of the 
accuracy of our approach.  

\begin{figure*}[hbtp]
\begin{center}
\begin{tabular}{cc}
\textsf{(A)} \resizebox{0.45\textwidth}{!}{\includegraphics{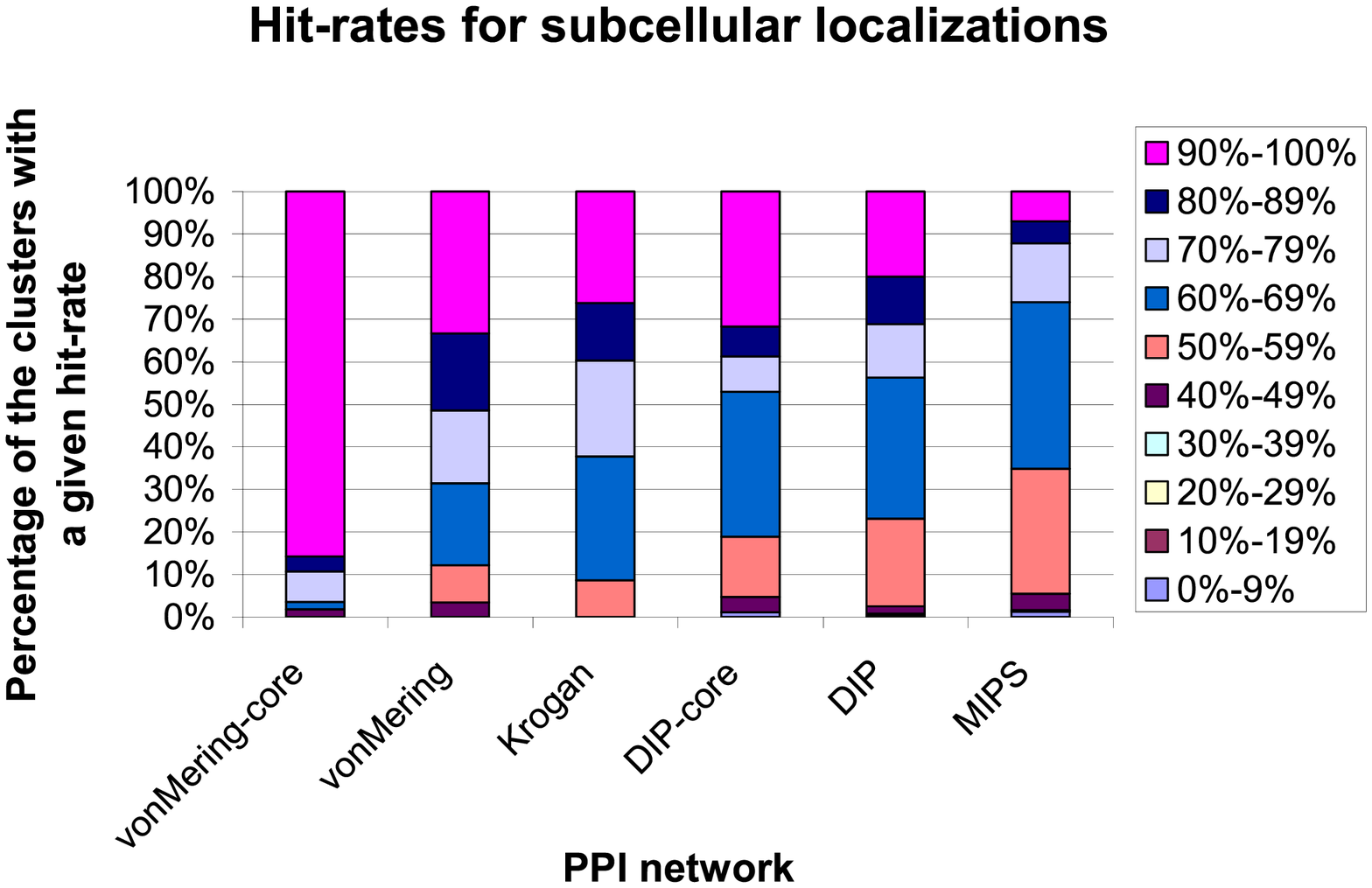}} &
\textsf{(B)} \resizebox{0.45\textwidth}{!}{\includegraphics{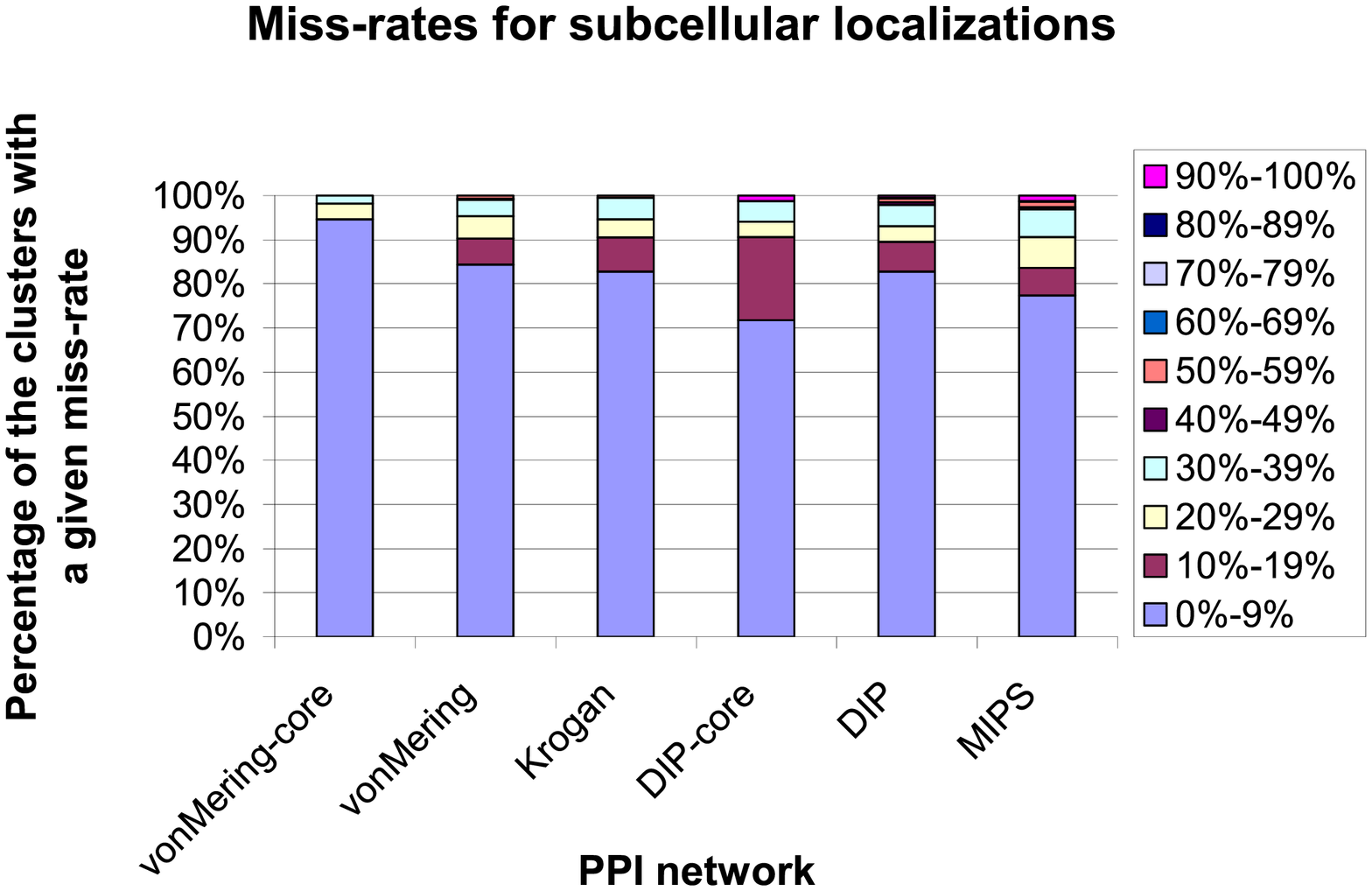}} \\
\resizebox{0.1\textwidth}{!}{\includegraphics{./figures/temp.ps}} \\
\resizebox{0.1\textwidth}{!}{\includegraphics{./figures/temp.ps}} \\
\textsf{(C)} \resizebox{0.45\textwidth}{!}{\includegraphics{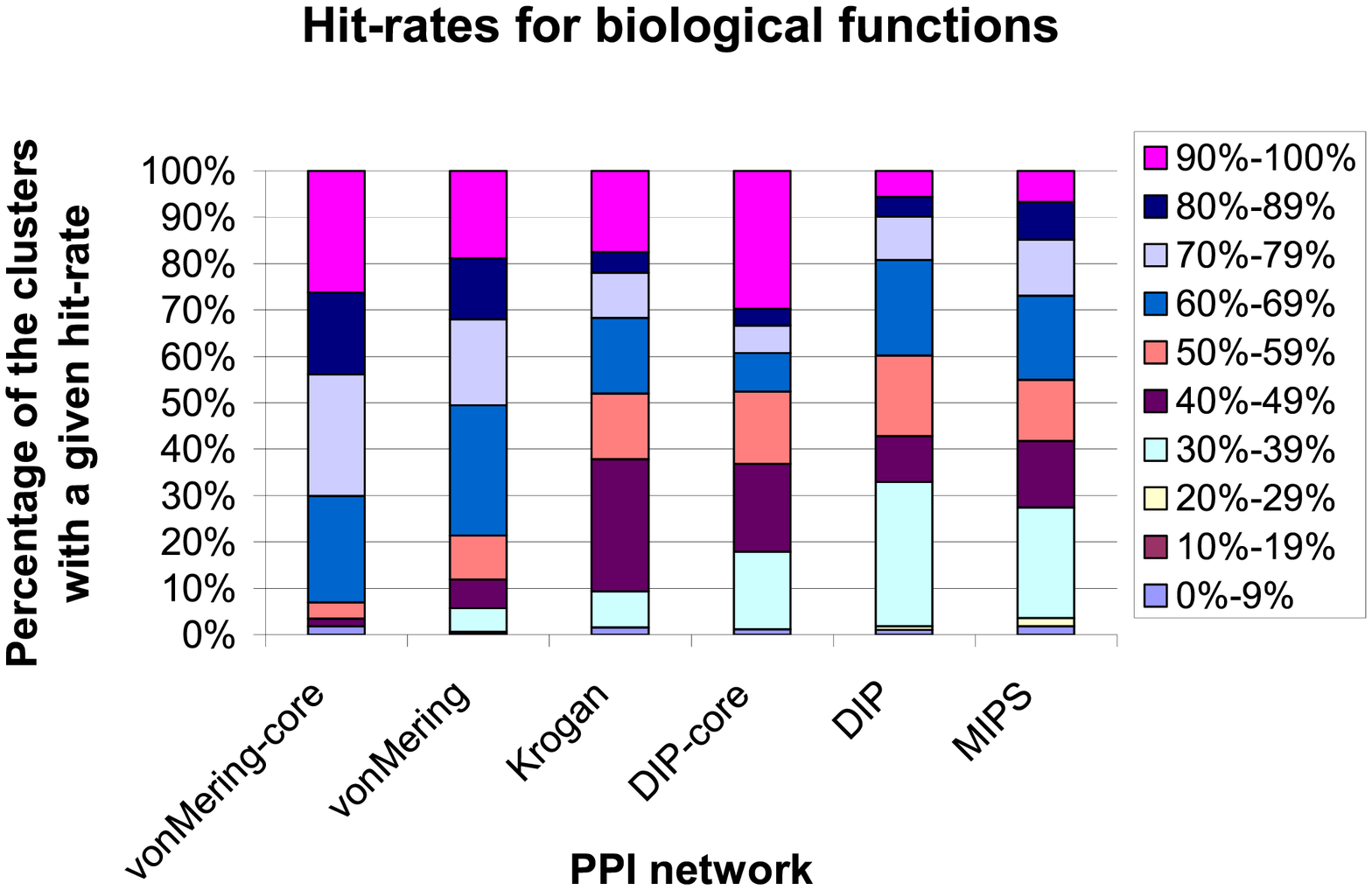}} &
\textsf{(D)} \resizebox{0.45\textwidth}{!}{\includegraphics{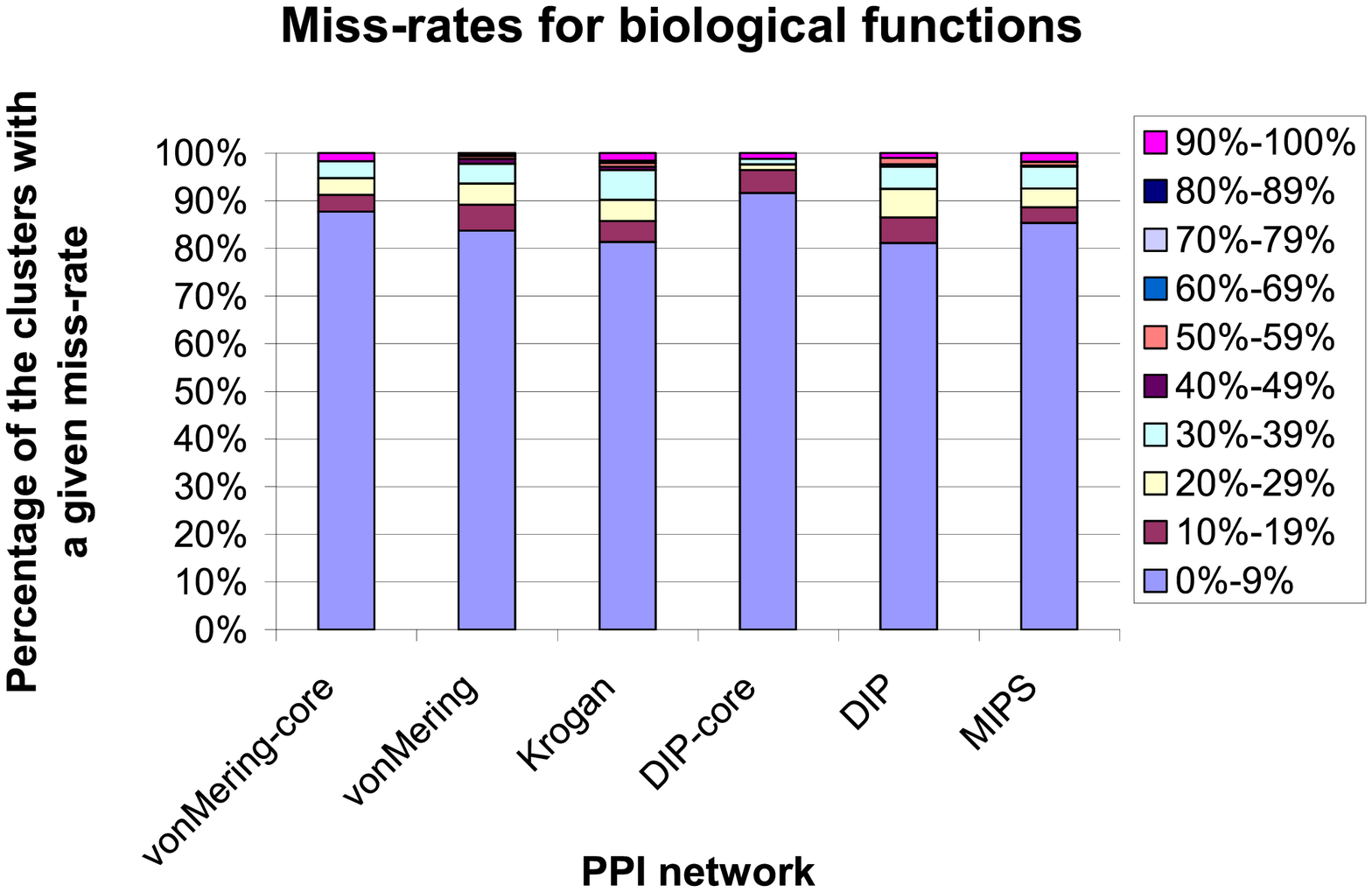}} \\
\resizebox{0.2\textwidth}{!}{\includegraphics{./figures/temp.ps}} \\
\resizebox{0.2\textwidth}{!}{\includegraphics{./figures/temp.ps}} \\
\textsf{(E)} \resizebox{0.45\textwidth}{!}{\includegraphics{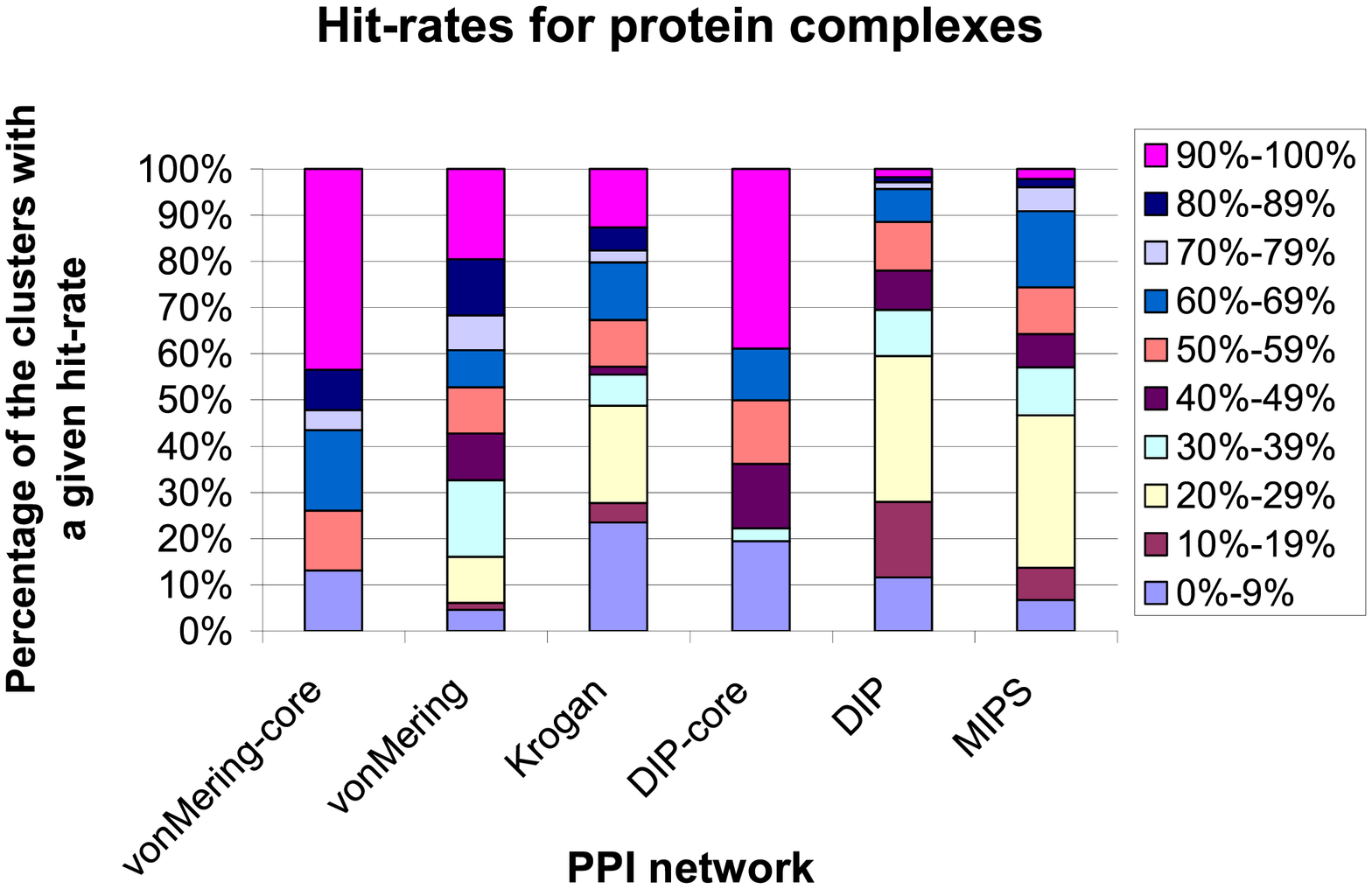}} &
\textsf{(F)} \resizebox{0.45\textwidth}{!}{\includegraphics{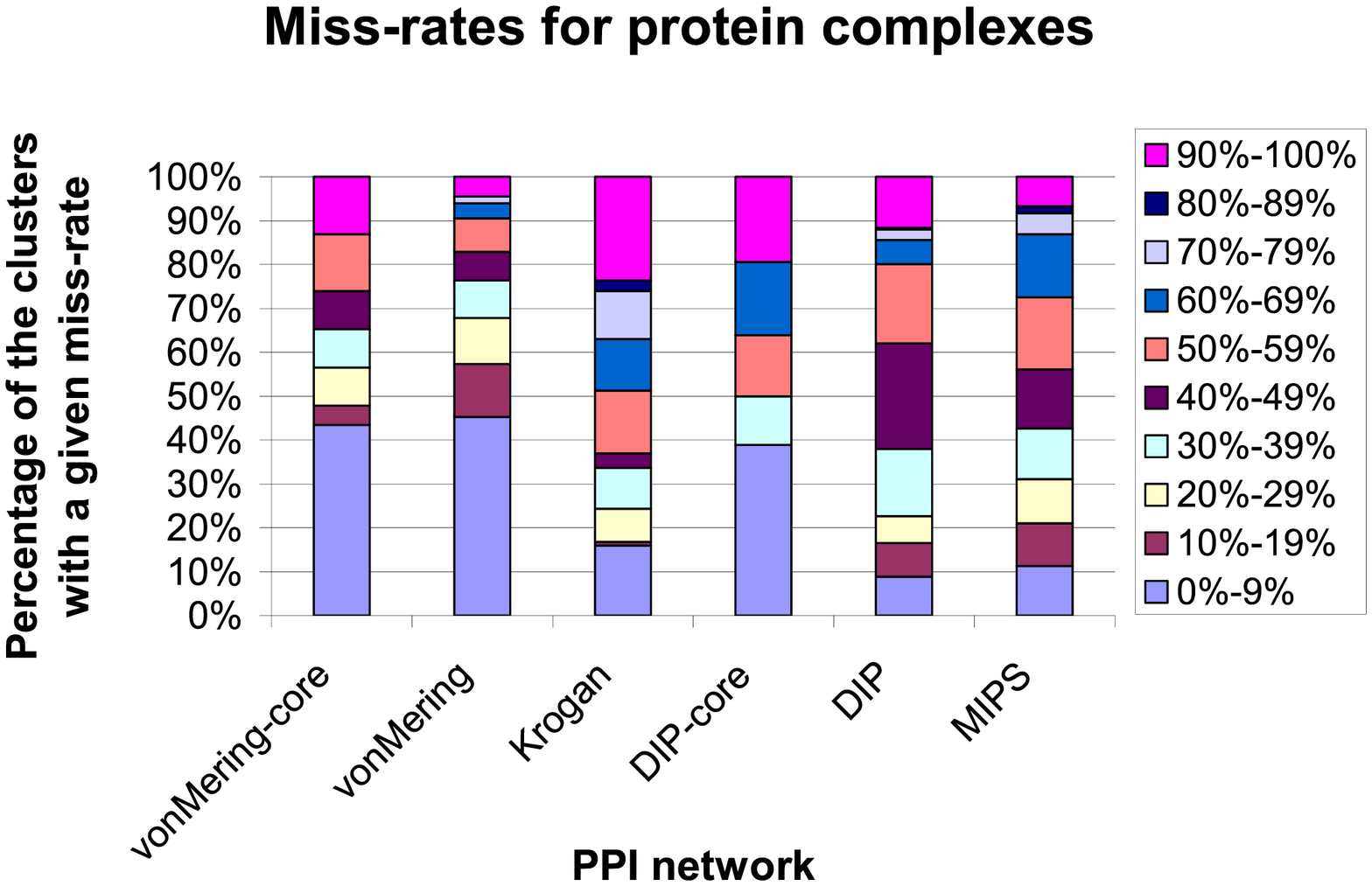}} \\
\end{tabular}
\caption{
The results of applying our method to the six yeast PPI networks 
(vonMering-core, vonMering, Krogan, DIP-core, DIP, and MIPS) 
and the three protein properties (subcellular localizations, 
biological functions, and protein complexes) in
accordance with the flexible scheme:
(A) hit-rates for subcellular localizations;
(B) miss-rates for subcellular localizations;
(C) hit-rates for biological functions;
(D) miss-rates for biological functions;
(E) hit-rates for protein complexes;
(F) miss-rates for protein complexes.
}
\label{fig:sum_res}
\end{center}
\end{figure*}

For each of the six yeast PPI networks, the three yeast
protein properties, and the two schemes, we measure the number of
clusters (out of the total number of clusters in a network) having 
given hit- and miss-rates.  We bin the hit- and miss-rates in
increments of 10\%.
The results for the flexible scheme are presented in
Figure \ref{fig:sum_res}. 
%
%
For subcellular localizations, 
in vonMering-core network, 86\% of
the clusters have hit-rate above 90\%; for the
remaining five networks, 65\% of clusters have hit-rates above 60\%
(Figure \ref{fig:sum_res} A). For all networks, miss-rates for 72\% of
clusters are bellow 10\% (Figure \ref{fig:sum_res} B). 
Similarly, for biological functions, the miss-rates in all six
networks are under 10\% for 81\% of the clusters
(Figure \ref{fig:sum_res} D). 
The hit-rates for biological functions are above 60\% for 79\% of
the clusters in both von Mering networks; in the remaining four
networks, 57\% of the clusters have hit-rates above 50\%
(Figure \ref{fig:sum_res} C). 
Finally, for protein complexes, 
47\% clusters in vonMering-core, vonMering, and DIP-core networks have hit-rates above 60\%, 
36\% of clusters in Krogan and MIPS networks have hit-rates above 50\%,
and 30\% of clusters in DIP network have hit-rates above 40\% 
(Figure \ref{fig:sum_res} E). 
Miss-rates for protein complexes are bellow 10\% for 39\% of
the clusters in both von Mering networks and in DIP-core network; in the remaining three
networks, 33\% of the clusters have miss-rates bellow 39\% (Figure \ref{fig:sum_res} F). 
%


\begin{figure*}[hbtp]
\begin{center}
\begin{tabular}{cc}
\textsf{(A)} \resizebox{0.38\textwidth}{!}{\includegraphics{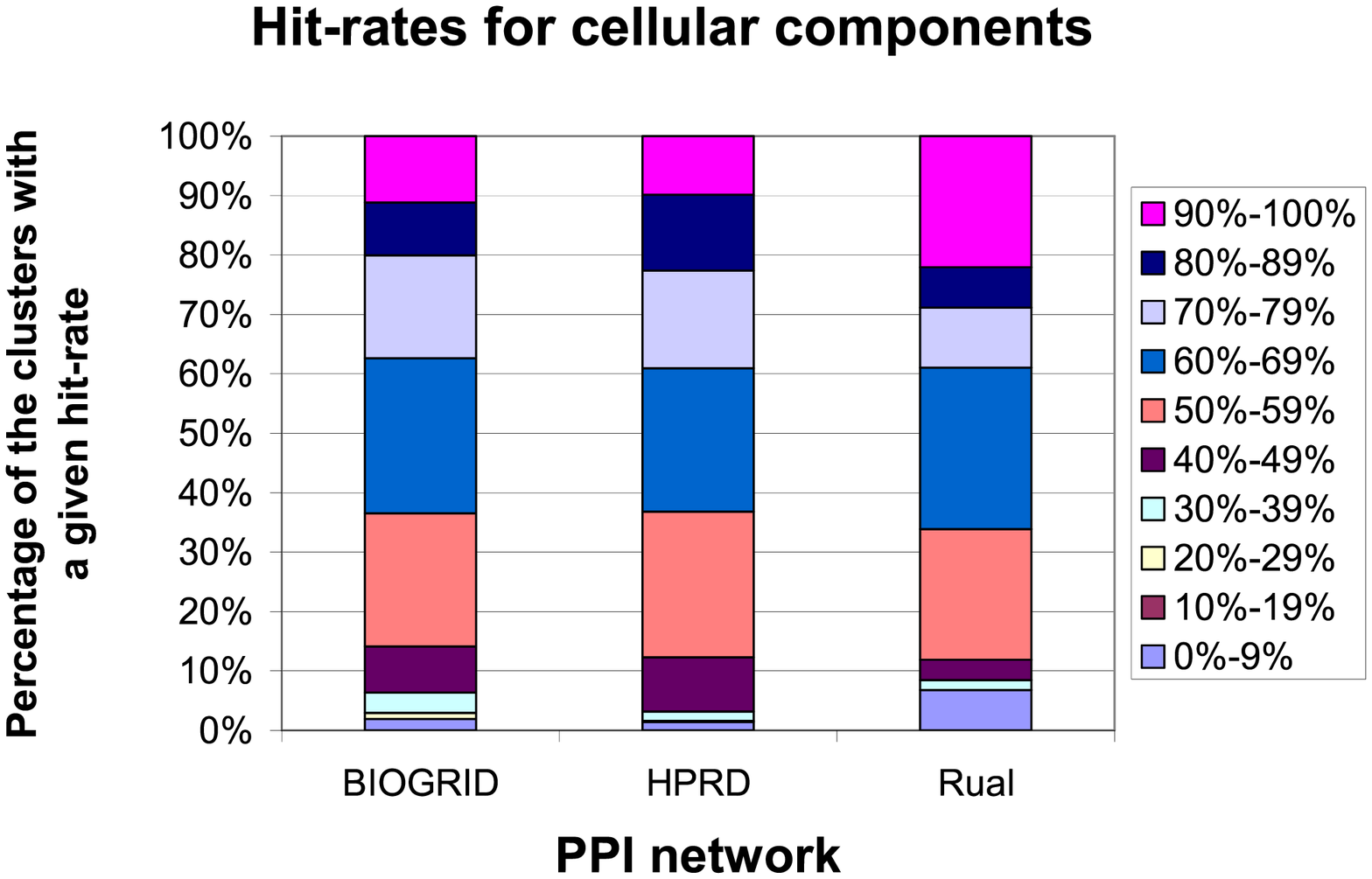}} &
\textsf{(B)} \resizebox{0.38\textwidth}{!}{\includegraphics{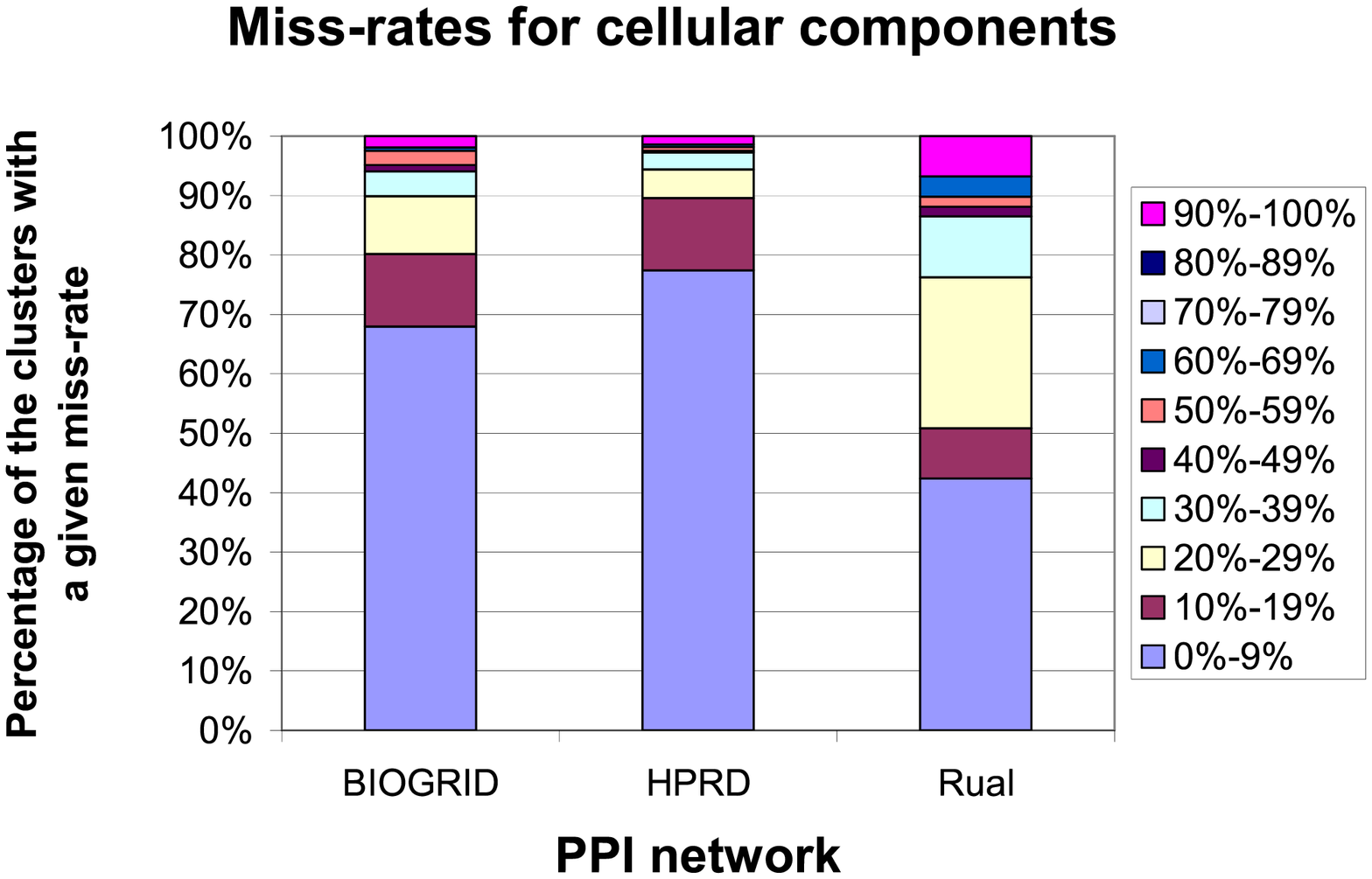}} \\
\resizebox{0.1\textwidth}{!}{\includegraphics{./figures/temp.ps}} \\
\resizebox{0.1\textwidth}{!}{\includegraphics{./figures/temp.ps}} \\
\textsf{(C)} \resizebox{0.38\textwidth}{!}{\includegraphics{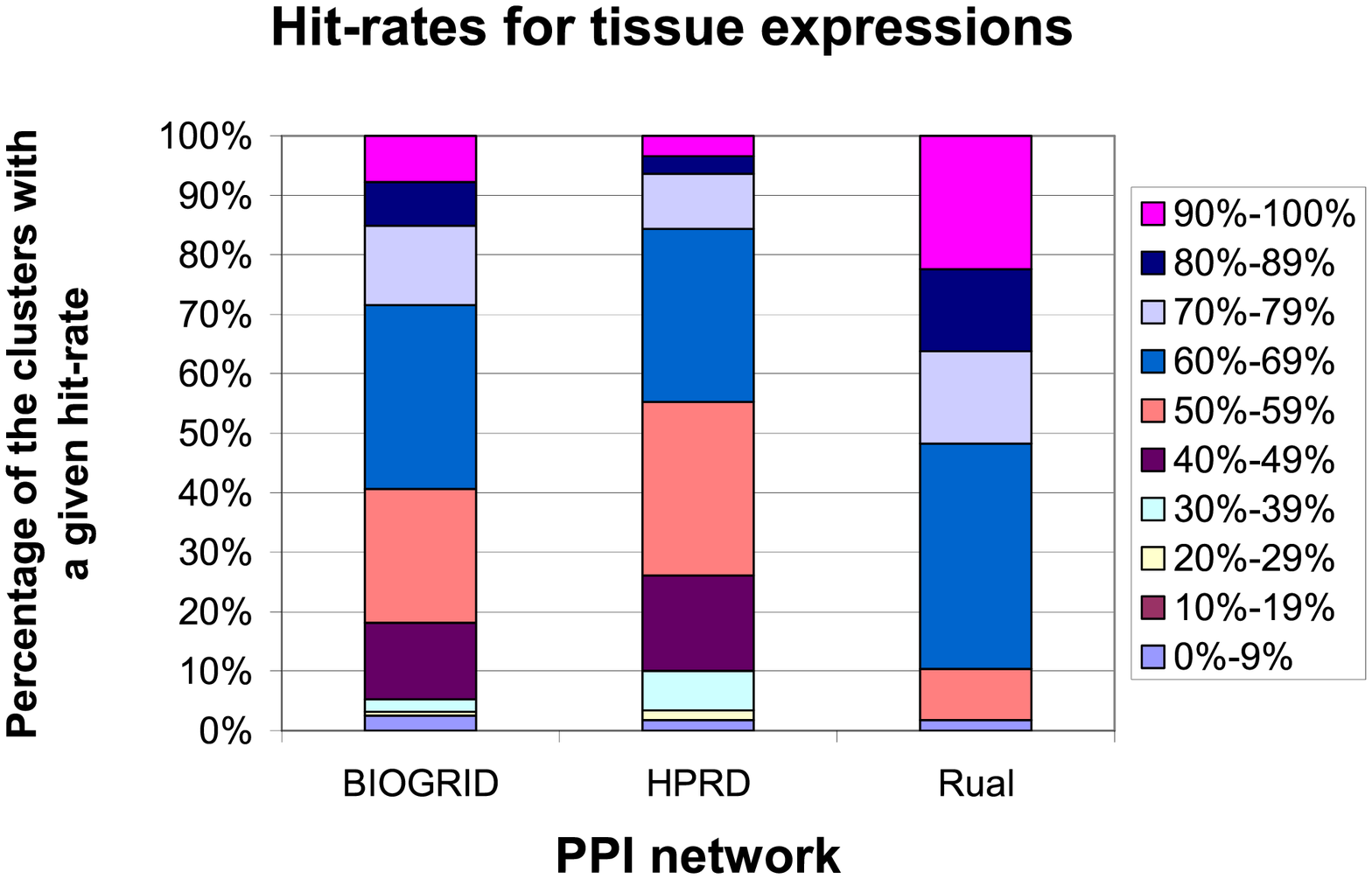}} &
\textsf{(D)} \resizebox{0.38\textwidth}{!}{\includegraphics{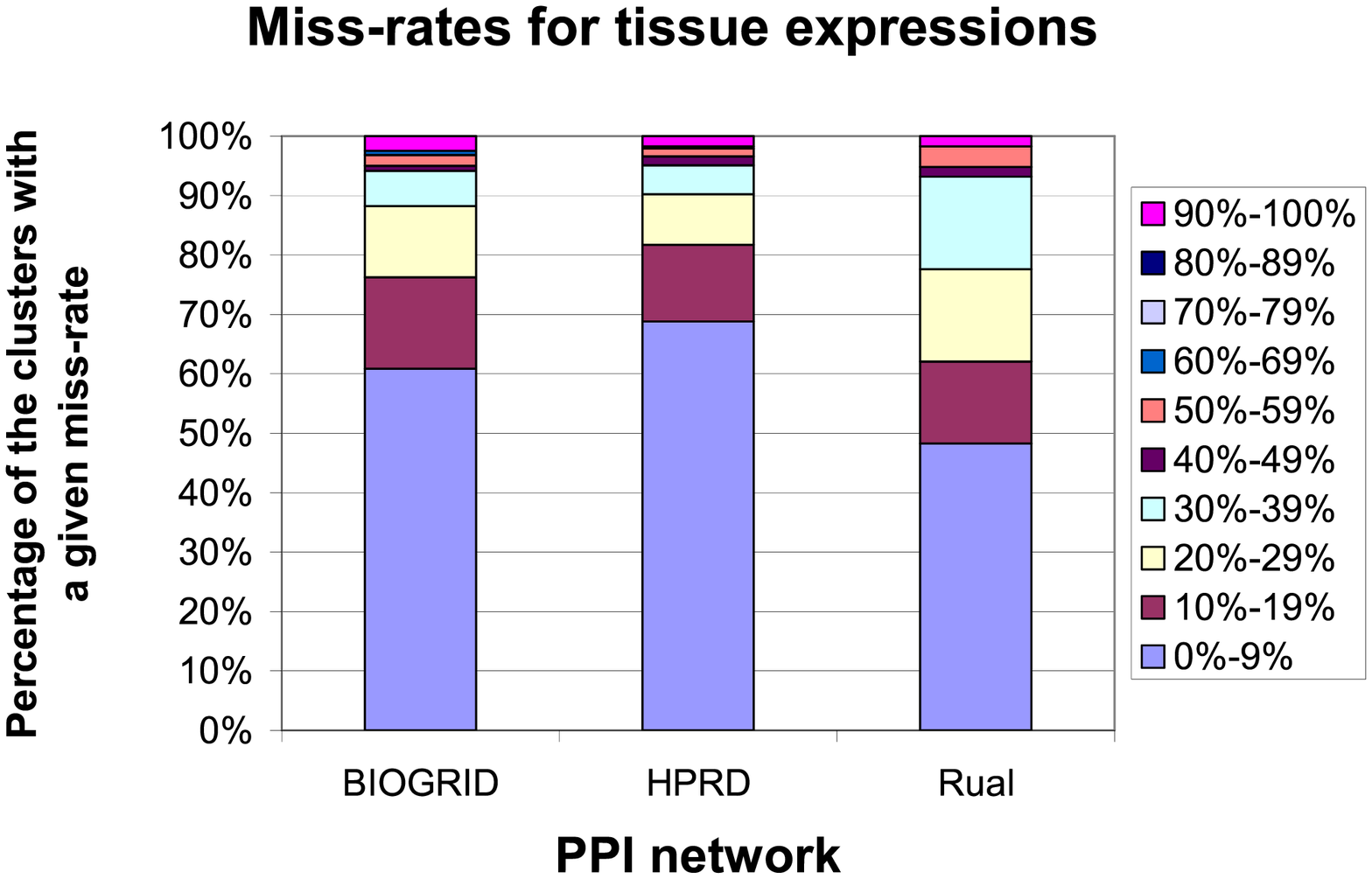}} \\
\resizebox{0.1\textwidth}{!}{\includegraphics{./figures/temp.ps}} \\
\resizebox{0.1\textwidth}{!}{\includegraphics{./figures/temp.ps}} \\
\textsf{(E)} \resizebox{0.38\textwidth}{!}{\includegraphics{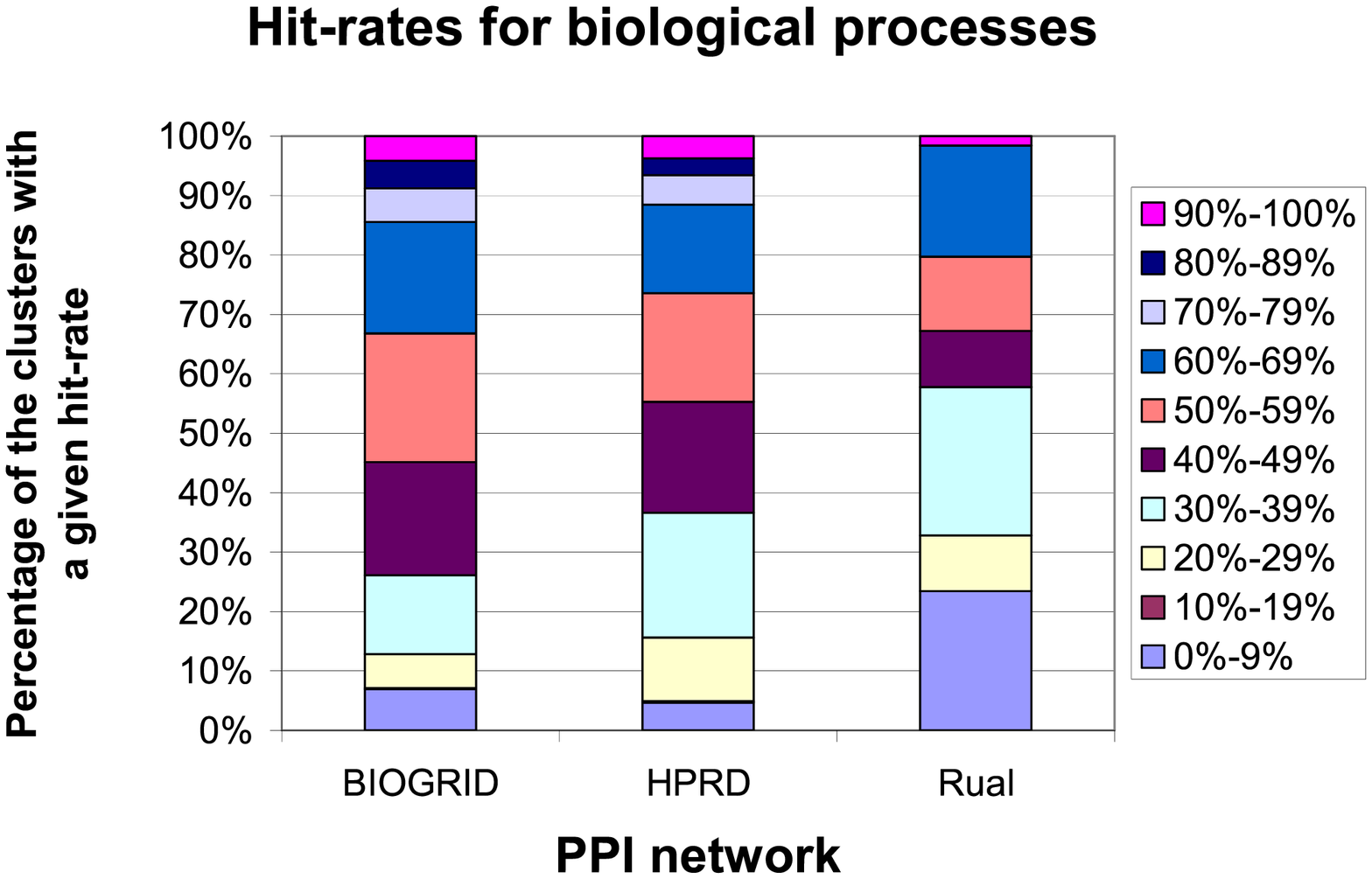}} &
\textsf{(F)} \resizebox{0.38\textwidth}{!}{\includegraphics{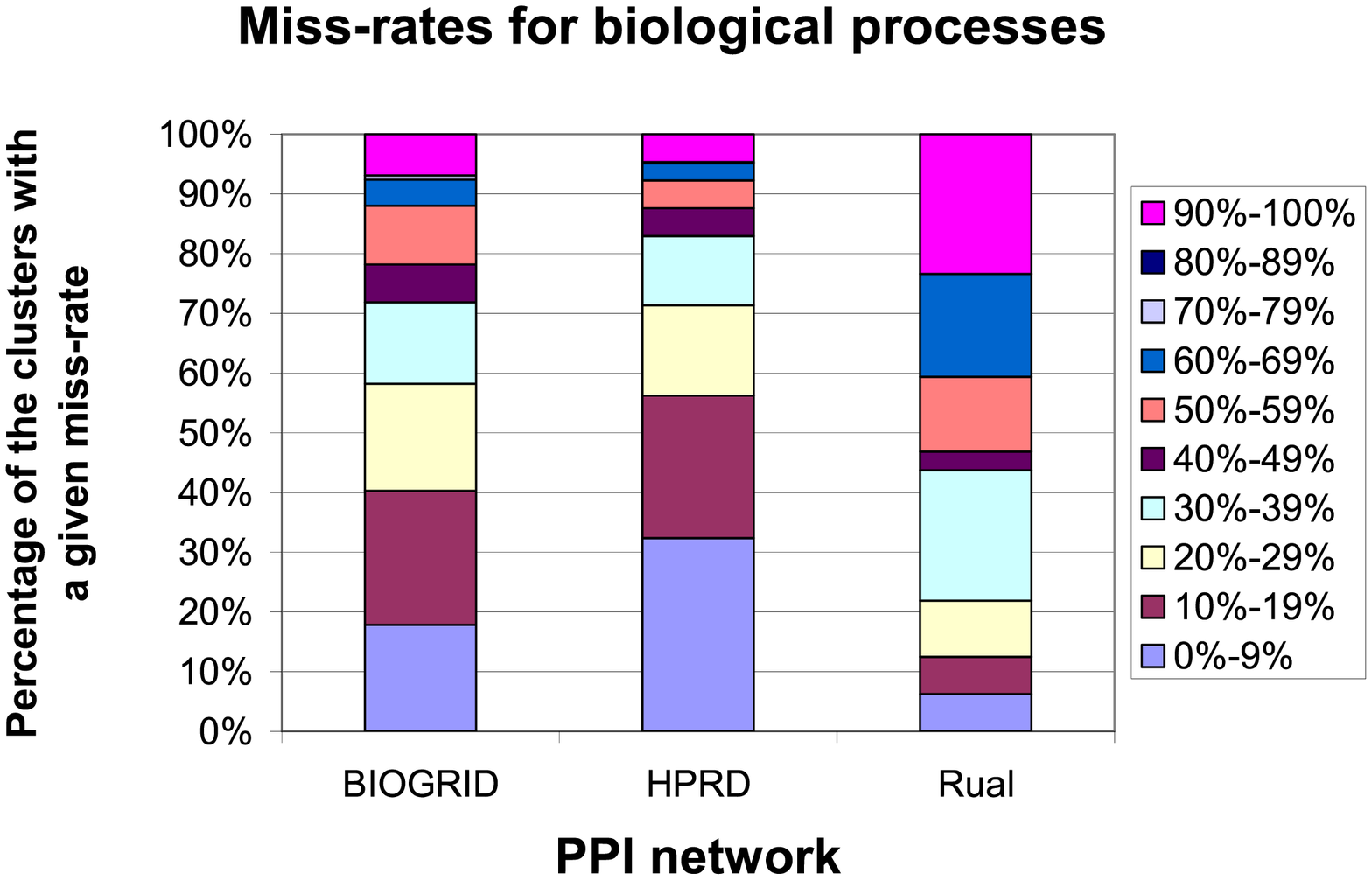}} \\
\end{tabular}
\caption{
The results of applying our method to the three human PPI networks 
(BIOGRID, HPRD, and Rual) 
and the three protein properties (cellular components, 
tissue expressions, and biological processes):
(A) hit-rates for cellular components;
(B) miss-rates for cellular components;
(C) hit-rates for tissue expressions;
(D) miss-rates for tissue expressions;
(E) hit-rates for biological processes;
(F) miss-rates for biological processes.
}
\label{fig:sum_res2}
\end{center}
\end{figure*}

Similarly, for each of the three human PPI networks and their three
protein properties that we analyzed, we measure the number of
clusters (out of the total number of clusters in a network) having 
given hit- and miss-rates.  
The results are presented in Figure \ref{fig:sum_res2}. 
For cellular components, in all three human PPI networks, 86\% of the clusters have hit-rates above 50\%
(Figure \ref{fig:sum_res2} A).  Miss-rates for 68\% of clusters in
BIOGRID and HPRD networks are bellow 10\%, while in Rual
network 76\% of clusters have miss-rates bellow 29\%
(Figure \ref{fig:sum_res2} B).  Similarly, for tissue expressions,
hit-rates are above 50\% for 74\% of clusters in BIOGRID and HPRD
networks, and for 98\% of clusters in Rual network, respectively
(Figure \ref{fig:sum_res2} C). 
Miss-rates are lower than 10\% for 61\%
of clusters in BIOGRID and HPRD networks, and for 48\% of clusters in
Rual network, respectively 
(Figure \ref{fig:sum_res2} D). 
Finally, for biological processes, hit-rates are above 50\% for 55\% of
clusters in BIOGRID network, for 45\% of clusters in HPRD network, and
for 33\% of clusters in Rual network, respectively.
(Figure \ref{fig:sum_res2} E). 
Miss-rates are bellow 29\% for 58\% of
the clusters in BIOGRID network and for 71\% of the clusters in HPRD network; 
in Rual network, 44\% of the clusters have miss-rates bellow 39\% (Figure \ref{fig:sum_res2} F). 


To evaluate the effect of noise in PPI networks to the accuracy of our method, 
we compare the results 
for the high-confidence vonMering-core network and the lower-confidence vonMering network (Figure \ref{fig:sum_res}). 
As expected, clusters in the more noisy network have lower hit-rates compared to the high-confidence network. 
However, low miss-rates are still preserved in clusters of both networks
for all three protein properties, 
indicating the robustness of our method to noise present in PPI networks.

Thus far, we demonstrated that our method identifies groups of nodes
in PPI networks having common protein properties.  Our technique can
also be applied to predict protein properties of yet unclassified
proteins by forming a cluster of proteins that are similar to the
unclassified protein of interest and assigning it the most common
properties of the classified proteins in the cluster. We do this for
all 115 functionally unclassified yeast proteins from MIPS that have
degrees higher than four in any of the six yeast PPI networks that we
analyzed.  In Tables \ref{tab:T1} and \ref{tab:T2}, we present the predicted
functions for proteins with prediction hit-rates of 50\% or higher
according to the strict and the flexible scheme, respectively.  The
full data set with functional prediction hit-rates lower than 50\%
is available upon request.
%
%
Note that a yeast protein can belong to more than one yeast PPI network that we analyzed.
%
%
Thus, biological functions that such proteins perform can be predicted from
clusters derived from different yeast PPI networks.  We observed an overlap
of the predicted protein functions obtained from multiple
PPI networks for the same organism, additionally
verifying the correctness of our method. 
Furthermore, there exists overlap between our protein function
predictions and those of others \citep{Vazquez2003}. 
Finally, we successfully predict the functional category of
PWP1 protein that is still functionally uncharacterized in MIPS, but
is characterized in SGD
\citep{SGD} as being involved in rRNA processing.

\begin{table*}[hbtp]
\begin{center}
\begin{tabular}{cc}
\resizebox{0.985\textwidth}{!}{\includegraphics{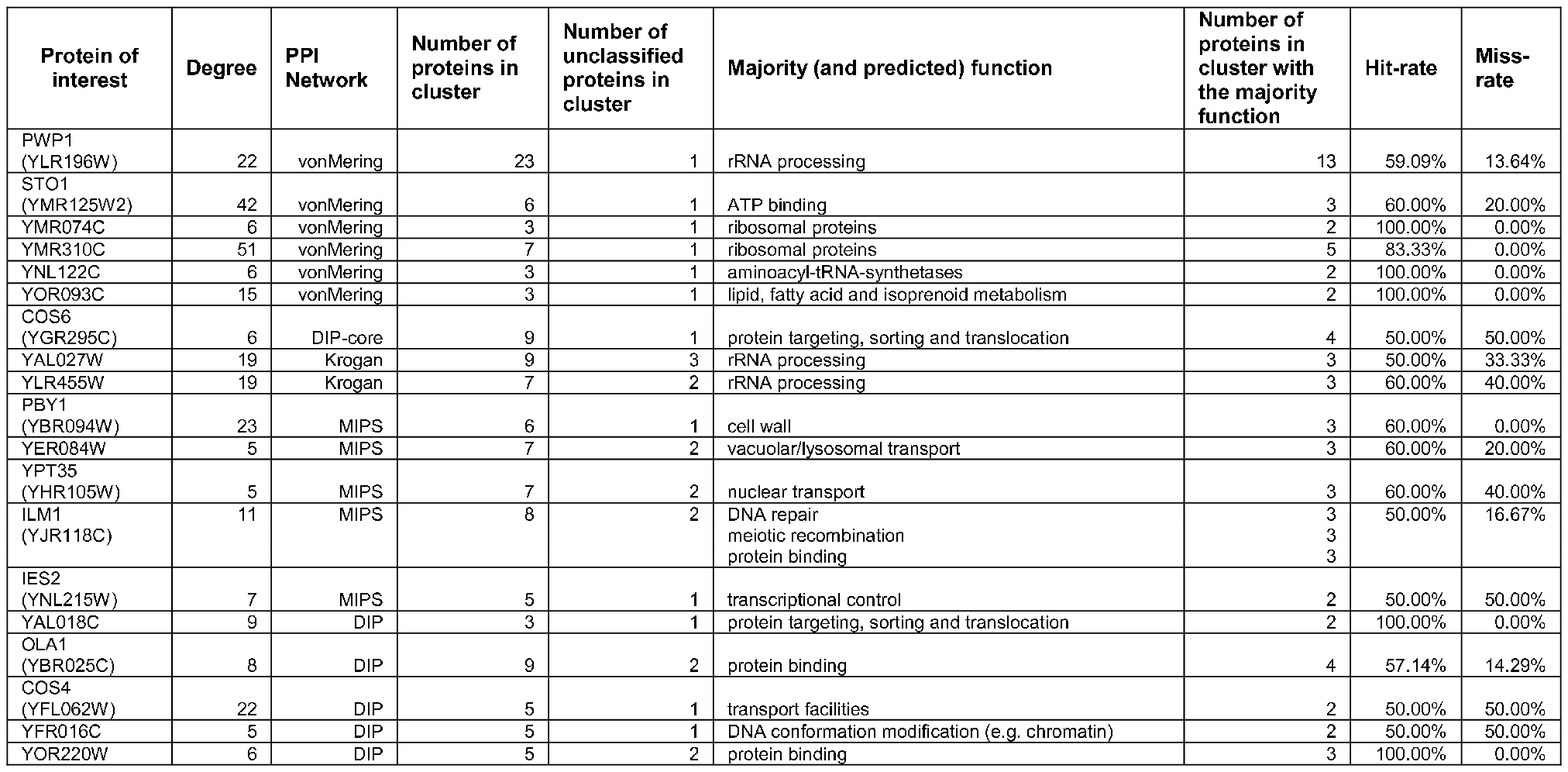}} &
\end{tabular}
\caption{Predicted functions with prediction
hit-rates of 50\% or higher according to the strict scheme for yeast
proteins that are unannotated in MIPS and that have degrees higher
than four in any of the six yeast PPI networks.  The column denoted by
``Protein of interest'' contains a protein of interest for which the
function is predicted. The column denoted by ``Degree'' contains the
degree of a given protein in the corresponding PPI network. The column
denoted by ``PPI Network'' contains the PPI network from which the
protein function was derived. The column denoted by ``Number of
proteins in cluster'' contains the total number of proteins in the
cluster, including the protein of interest. The column denoted by
``Number of unclassified proteins in cluster'' contains the number of
functionally unclassified proteins in a given cluster, including the
protein of interest. The column denoted by ``Majority (and predicted)
function'' contains the common functions amongst at least 50\% proteins
in the cluster that are also predicted functions for the protein of
interest. The column denoted by ``Number of proteins in cluster with
the majority function'' contains the number of nodes in the cluster
with the majority function. The column denoted by ``Hit-rate'' contains
the percentage of the total number of proteins in the cluster with the
majority function; only the maximum hit-rate is reported for a protein
of interest. Finally, the column denoted by ``Miss-rate'' contains the
percentage of annotated nodes in the cluster that do not have a common
function with any other annotated node in the cluster. 
}
\label{tab:T1}
\end{center}
\end{table*}

\begin{table*}[hbtp]
\begin{center}
\begin{tabular}{cc}
\resizebox{0.985\textwidth}{!}{\includegraphics{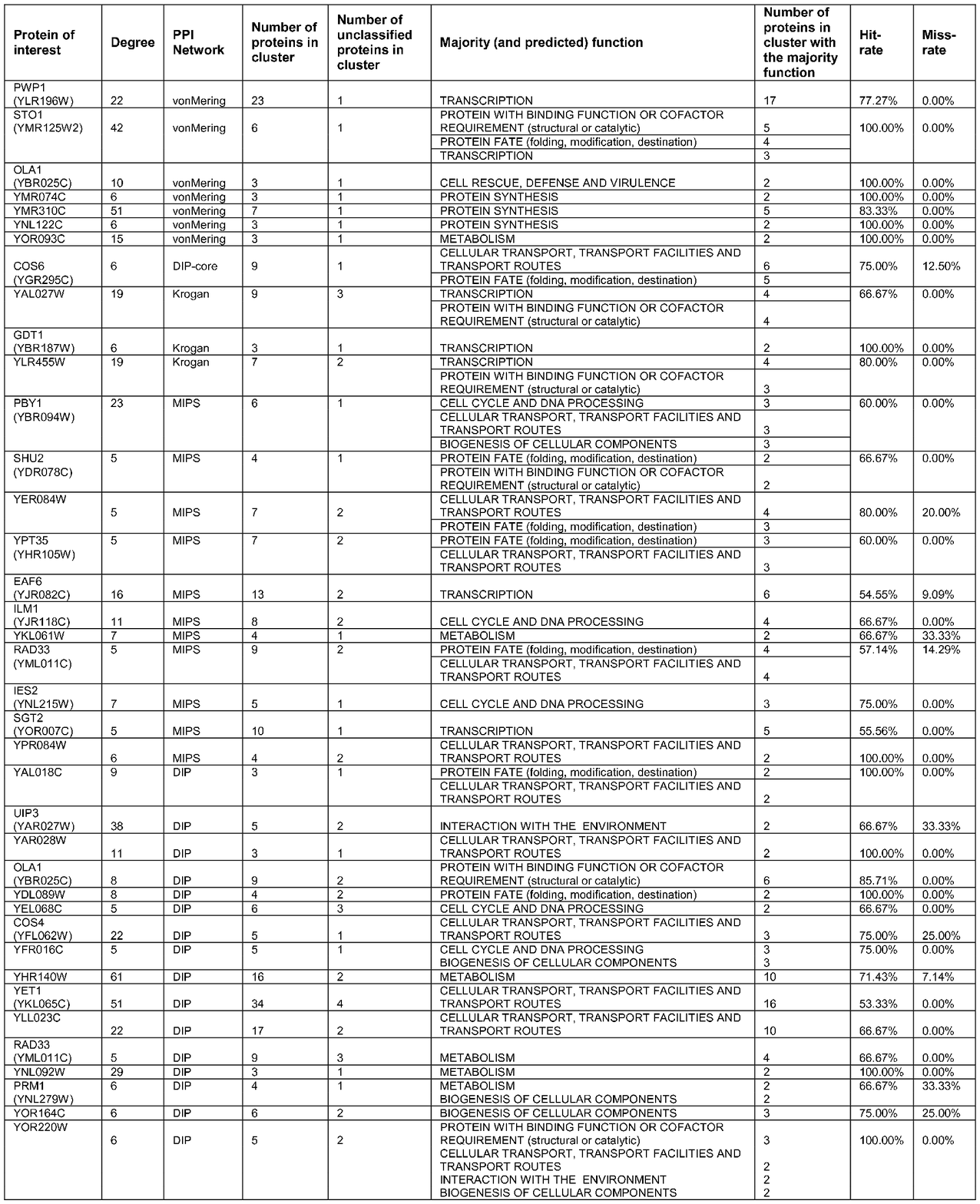}} &
\end{tabular}
\caption{
Predicted functions with prediction
hit-rates higher than 50\% according to the flexible scheme for yeast
proteins that are unannotated in MIPS and that have degrees higher
than four in any of the six yeast PPI networks.  The columns have
the same meaning as in Table \ref{tab:T1}. 
}
\label{tab:T2}
\end{center}
\end{table*}

To our knowledge, this is the first study that relates the PPI network
structure to all of the following: protein complexes, biological
functions, and subcellular localizations for yeast, 
and cellular components, tissue expressions, and 
biological processes for human.
Starting with the topology of PPI networks of different organisms that are of different 
sizes and are originating from a wide spectrum
of small-scale and high-throughput PPI detection techniques, our
method identifies clusters of nodes sharing common protein properties.
Our method accurately uncovers groups of nodes belonging to the same
protein complexes in the vonMering-core network:
44\% of clusters have 100\% hit-rate according to the flexible scheme. 
This additionally validates our method, since PPIs in this network are obtained mainly by 
TAP \citep{Rigaut99,Gavin02}
and HMS-PCI \citep{Ho2002}, 
which are known to favor protein complexes.

Our node similarity measure is highly constraining, since we take into
account not only a node's degree, but also additional 72 ``graphlet
degrees'' (see section \ref{methods}).  Since the number of graphlets
on $n$ nodes increases exponentially with $n$, we use 2-5-node
graphlets (see Figure \ref{fig:graphlet_orbits}). However, our method is
easily extendible to include larger graphlets, but this would increase
the computational complexity; the complexity is currently
$O(|V|^5)$ for a graph $G(V,E)$, since we search for graphlets with up
to $5$ nodes.  Nonetheless, since our algorithm is ``embarrassingly
parallel'' (i.e., can easily be distributed over a cluster of
machines), extending it to larger graphlets is feasible.  In addition
to the design of the
signature similarity measure as a number in [0, 1], this makes
our technique usable for larger networks.

\section{Conclusion}

We present a new graph theoretic method for detecting the relationship
between local topology and function in real-world networks.  We apply
it to proteome-scale PPI networks and demonstrate the link between the
topology of a protein's neighborhood in the network and its membership
in protein complexes, functional groups, and subcellular compartments for yeast, 
and in cellular components, tissue expressions, and biological processes for human.
Additionally, we demonstrate that our method can be used to predict
biological function of uncharacterized proteins.  Moreover, the method
can be applied to different types of biological and other real-world
networks and give insight into complex biological mechanisms and 
guidelines for future experimental research.

\section*{Funding}
This project was supported by the NSF CAREER IIS-0644424 grant.

\section*{Acknowledgement}
We thank the Institute for Genomics and Bioinformatics (IGB) at UC
Irvine for providing computing resources. 

\bibliographystyle{chicago}
\bibliography{all}

\end{document}